\documentclass[epj]{webofc}
\usepackage[utf8]{inputenc}
\usepackage[varg]{txfonts}   
\usepackage{booktabs}
\usepackage{xcolor}
\definecolor{darkred}{rgb}{0.4,0.0,0.0}
\definecolor{darkgreen}{rgb}{0.0,0.4,0.0}
\definecolor{darkblue}{rgb}{0.0,0.0,0.4}
\usepackage[bookmarks,linktocpage,colorlinks,
    linkcolor = darkred,
    urlcolor  = darkblue,
    citecolor = darkgreen]{hyperref}
\usepackage{subfigure}
\wocname{EPJ Web of Conferences}
\woctitle{Lattice2017}
\begin{document}
%
\selectlanguage{english}
\title{%
  SU(3) breaking and the pseudo-scalar spectrum in multi-taste QCD
}
\author{%
  \firstname{Michael} \lastname{Creutz}\inst{1,3}\fnsep\thanks
            {Acknowledges partial travel support
under
contract number DE-AC02-98CH10886 with the U.S.~Department of Energy.
Accordingly, the U.S. Government retains a non-exclusive, royalty-free
license to publish or reproduce the published form of this
contribution, or allow others to do so, for U.S.~Government purposes.}
            \fnsep\thanks{Speaker, \email{mike@latticeguy.net}
              }
}
\institute{%
  Physics Department, Brookhaven National Lab,
  Upton, NY 11973, USA
  }
\abstract{Using the Sigma model to explore the lowest order
  pseudo-scalar spectrum with SU(3) breaking, this talk considers an
  additional exact "taste" symmetry to mimic species doubling. Rooting
  replicas of a valid approach such as Wilson fermions reproduces the
  desired physical spectrum. In contrast, extra symmetries of the
  rooted staggered approach leave spurious states and a flavor
  dependent taste multiplicity.  }
\maketitle
\section{Introduction}\label{intro}

Despite over 10 years of controversy, the
rooting trick\cite{Bernard:2006qt} proposed for reducing the taste
structure of staggered
fermions\cite{Kogut:1974ag,Susskind:1976jm,Sharatchandra:1981si} is
not valid\cite{Creutz:2007yg,Creutz:2007rk}. The issue is that the
exact chiral symmetries inherent in the formulation are incompatible
with the well known chiral anomaly.  

As the algorithm remains popular and these issues are somewhat subtle,
this talk concentrates on two of the more blatant issues that arise
when the $SU(3)$ flavor symmetry is broken via unequal quark masses.
First, different elements of the usual $SU(3)$ pseudo-scalar octet
appear with different taste degeneracies.  Second, a large number of
spurious pseudo-scalars appear far from any physical
particle masses.

Section \ref{sec-1} reviews the standard sigma model picture of the
pseudo-scalar mass dependence on the underlying quark masses.  Section
\ref{sec-2} augments that argument to include the taste degeneracies
inherent in the staggered approach.  Section \ref{sec-3} briefly
discusses how rooting can work with multiple copies, or replicas, of a
valid lattice fermion formulation, such as Wilson
fermions\cite{Wilson:1975id}.  Section \ref{sec-4} contrasts that
argument with staggered fermions, which are not replicas of equivalent
fermions.  Here the inherent propagator structure forces the spurious
states of the unrooted theory to survive.  Finally Section \ref{sec-5}
reiterates the two main points mentioned above.

\section{Quark masses and the pseudo-scalar spectrum}\label{sec-1}

Consider three flavor QCD and the usual pseudo-scalar octet
\begin{equation}
\begin{matrix}
& K_0 & & K_+ & \cr
\cr
\pi_- && \pi_0,\eta &&\pi_+\cr
\cr
& K_- & & \overline K_0 & \cr 
  \end{matrix}.
\end{equation}  
This section reviews the lowest-order non-linear sigma model
prediction for the mass dependence of these particles on the
underlying quark masses.  For simplicity, ignore the $\eta^\prime$ on
the grounds that it acquires a large mass through the anomaly.

Begin with the usual picture of spontaneous chiral symmetry breaking
giving a quark condensate
\begin{equation}
\langle\overline\psi\psi\rangle=v
\end{equation}
where $\psi$ denotes the quark fields.  The non-linear sigma model is
an effective theory for fluctuations around this expectation value
\begin{equation}
\overline\psi_L^j\psi_R^k \sim v\ \Sigma^{jk}. 
\end{equation}
Here the roman superscripts denote the quark flavors, which run from 1
to 3, or equivalently lie in $\{ u,d,s\}$.  Ignoring radial
fluctuations, the matrix $\Sigma$ is taken to lie in the group
$SU(3)$.

Introduce the eight Gell-Mann matrices $\lambda_\alpha$ normalized
\begin{equation}
{\rm Tr} \lambda_\alpha \lambda_\beta =
  2\delta_{\alpha\beta}.
\end{equation}
Contact with the pseudo-scalar fields follows from
\begin{equation}
\Sigma=\exp(i\pi_\alpha \lambda_\alpha/f_\pi).
\end{equation}
Here $f_\pi$ is a phenomenological constant of about 93 MeV.

The kinetic term for the effective field $\Sigma$ takes the form
\begin{equation}
L_0={f_\pi^2\over 4}{\rm Tr}(\partial_\mu \Sigma^\dagger \partial_\mu \Sigma).
\end{equation}
Expanding this to second order in the pion fields gives their
effective kinetic term
\begin{equation}
L_0={\rm const}+{1\over 2} \partial_\mu\pi_\alpha \partial_\mu\pi_\alpha+\ldots
\end{equation}

\begin{figure}[thb] 
  \centering
  \includegraphics[width=.7\hsize,clip]{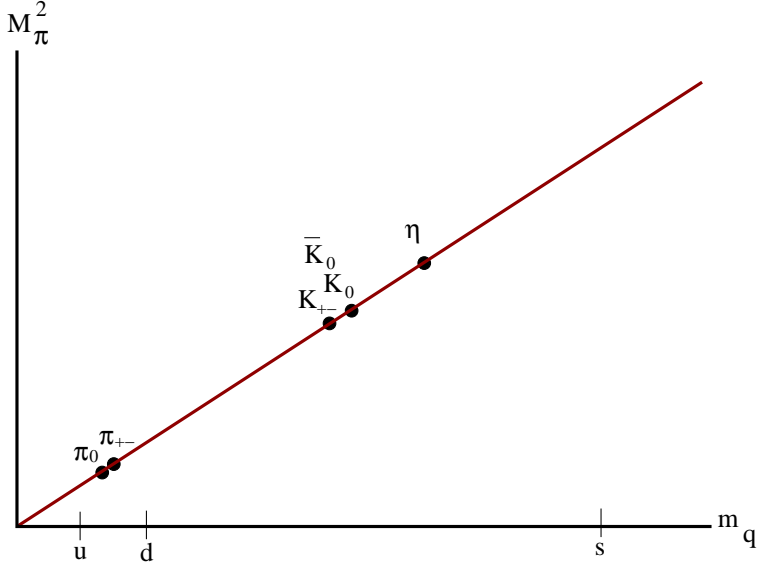}
  \caption{The pseudo-scalar spectrum for fixed quark masses as
    predicted by the effective sigma model.  The quark masses are
    indicated on the x axis.}
  \label{fig-1}
\end{figure}

When the quark masses vanish, the starting theory has two chiral
symmetries under the rotations
\begin{equation}
\begin{matrix}
&\psi_L\rightarrow \psi_L\ g_L\cr
&\psi_R\rightarrow \psi_R\ g_R\cr
\end{matrix}
  \end{equation}
where $g_L$ and $g_R$ are independent global elements of $SU(3)$.
In the effective theory this symmetry takes the form
\begin{equation}
  \Sigma\rightarrow g_L^\dagger\ \Sigma\ g_R.
\end{equation}
The spontaneous breaking of this symmetry gives the usual octet of
pseudo-scalars.

Quark masses break both the flavor $SU(3)$ and the above chiral
symmetries.\footnote{Electromagnetic effects are ignored here.}
For the effective theory add a mass term and
consider
\begin{equation}
L=L_0- {vf_\pi^2\over 4} {\rm Re\ Tr}(m\ \Sigma)
\end{equation}
with $m$ a 3 by 3 matrix.  Chiral rotations of form
$m\rightarrow g_R^\dagger\ m\ g_L$ allows $m$ to be put in diagonal
form
\begin{equation}
m=\begin{pmatrix} 
m_u & 0 & 0 \cr
0 & m_d & 0 \cr
0 & 0 & m_s \cr
\end{pmatrix}.
\end{equation}
Expanding $L$ to quadratic order in the pion fields gives
\begin{equation}
L={\rm const}+{1\over 2}\partial_\mu \pi_\alpha\partial_\mu \pi_\alpha
+{1\over 2}\pi_\alpha M_{\alpha\beta}\pi_\beta
  \end{equation}
where the 8 by 8 meson mass matrix $M$ takes the form
\begin{equation}
M_{\alpha\beta} =  {\rm Re\ Tr}\ \lambda_\alpha m \lambda_\beta.
  \end{equation}

From this it is elementary algebra to obtain the octet masses
\begin{equation}
  \begin{matrix}
    &M_{\pi_+}^2= \  M_{\pi_-}^2\propto {m_u+m_d \over 2}\hfill\cr
&M_{K_+}^2= \ M_{K_-}^2\propto  {m_u+m_s \over 2} \hfill\cr
&M_{K_0}^2= \ M_{\overline K_0}^2\propto { m_d+m_s \over 2} \hfill \cr
& M_{\pi_0}^2 \propto\  {1\over 3} \bigg(m_u+m_d+m_s
-\sqrt{m_u^2+m_d^2+m_s^2-m_um_d-m_um_s-m_dm_s}\bigg)\cr
&M_{\eta}^2 \propto \ {1\over 3} \bigg(m_u+m_d+m_s
+\sqrt{m_u^2+m_d^2+m_s^2-m_um_d-m_um_s-m_dm_s}\bigg)\cr
    \end{matrix}.
  \end{equation}
Note that this involves solving a quadratic equation for the
$\pi_0$ $\eta$ mixing arising because $M_{38}$ does not vanish.
This spectrum is qualitatively sketched in
Fig. \ref{fig-1}.

\section{Including taste degeneracy}\label{sec-2}

Motivated by the four tastes inherent with staggered fermions,
introduce a factor of $N_t=4$ degeneracy for each quark flavor.  This
leaves us with 12 distinct quark species\cite{Lee:1999zxa}. To keep
the algebra simple, assume that this ``taste'' symmetry is exact for
each of the original three ``flavors.''  Before the breaking of flavor
by the quark masses, the chiral symmetry becomes $SU(12)\otimes
SU(12)$.  Thus there should be $143=12^2-1$ pseudo-Goldstone bosons.
The 8 by 8 meson mass matrix becomes 143 by 143, which will now be
diagonalized.\footnote{As the goal is an eventual reduction to the
  normal 3 flavor theory, ignore the possibility of the confining
  theory reverting to a conformal one.}

This diagonalization is simplified since there are actually three
distinct $SU(4)$ taste groups, one for each of the flavors
$u,\ d,\ s$.  This makes it possible to classify the pseudo-scalar
mesons in terms of their representations under each of these groups.  The
relevant $SU(4)$ representations are the $1,4, \overline 4, 15$ in
analogy to the $SU(3)$ representations $1,3,\overline 3,8$.

The kaons and charged pions are particularly easy to treat.  Each
involves two distinct flavors, $q$ and $q^\prime$.  The mesons appear in the
representation $(4_q, \overline 4_{{\overline q}^\prime})$.  Thus they
form multiplets of 16 mesons each.  The meson masses squared are
proportional to the average of their constituent masses, {\em i.e.}
$M^2\propto {1\over 2}(m_q+m_{q^\prime})$.  With four kaons and two
charged pions, this accounts for $6\times16=96$ of our total 143
expected pseudo-Goldstone particles.

Now turn to the neutral mesons, those for which a quark is combined
with its anti-quark.  For this use the breakdown $
4\otimes \overline 4 \rightarrow 1 \oplus 15$.  For each flavor,
begin with a taste 15 plus a taste singlet.

Remarkably, the taste 15 combinations for each of the flavors cannot
mix.  This is because each flavor has its own taste group.  This gives
\begin{itemize}
  \item a taste 15 of {$i\overline u \gamma_5 u$}
states with $M^2\propto m_u$,
\item a taste 15 of {$i\overline d \gamma_5 d$}
states with $M^2\propto m_d$,
\item a taste 15 of {$i\overline s \gamma_5 s$}
states with $M^2\propto m_s$.
\end{itemize}
None of these has any analogue in the spectrum of Fig. \ref{fig-1}.

\begin{figure}[thb] 
  \centering
  \includegraphics[width=.7\hsize,clip]{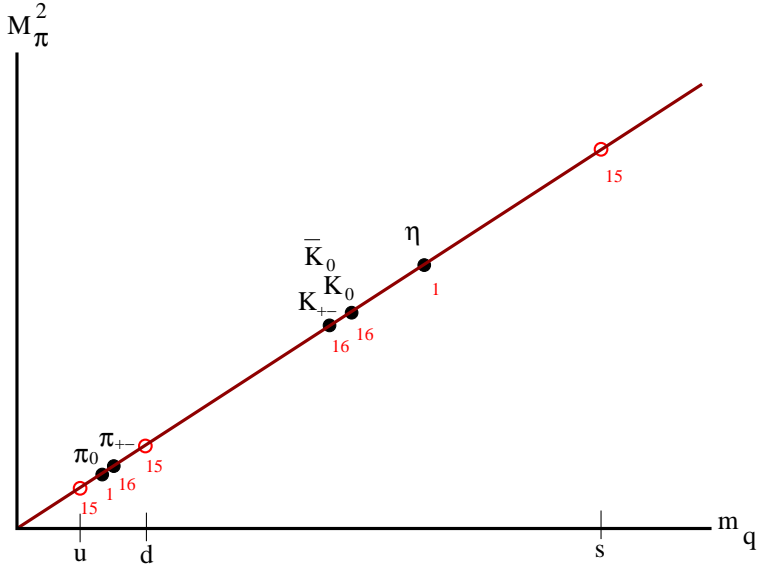}
  \caption{The pseudo-scalar spectrum for the multi-taste theory.  The
    degeneracies of the various states are indicated below the line.
    The 15-plets, denoted by open circles, do not appear in the single
    taste theory.  Note that the mesons with physical masses do not
    have a common taste multiplicity.}
  \label{fig-2}
\end{figure}

Finally consider the taste singlet combinations $\overline uu,\overline
dd, \overline ss$.  From these three, the flavor singlet combination is
the $\eta^\prime$.  As stated earlier, this is heavy and ignored here. The
remaining two combinations have the identical mixing matrix as
discussed earlier for the single taste theory.  These give rise to the
masses quoted earlier
\begin{equation}
  \begin{matrix}
& M_{\pi_0}^2 \propto\  {1\over 3} \bigg(m_u+m_d+m_s
-\sqrt{m_u^2+m_d^2+m_s^2-m_um_d-m_um_s-m_dm_s}\bigg)\cr
&M_{\eta}^2 \propto \ {1\over 3} \bigg(m_u+m_d+m_s
+\sqrt{m_u^2+m_d^2+m_s^2-m_um_d-m_um_s-m_dm_s}\bigg).\cr
    \end{matrix}
\end{equation}
This relation for the $\eta$ mass in the multi-taste theory appears
in Ref. \cite{Aubin:2003mg}.

All anticipated states are now identified, {\em i.e.}  $143=16\times
6+15\times 3+2$.  The resulting spectrum is sketched in
Fig. \ref{fig-2}.  The important points to note in comparing this with
Fig. \ref{fig-1} are that the taste degeneracies of the physical
states depend on which element of the octet one is observing, and
there are three multiplets of 15 particles each that have no
correspondence in the single taste theory.

\section{Rooting replicas}\label{sec-3}

Consider a theory with 4 replicas of a valid fermion formulation,
such as Wilson fermions.  A reduction of the four taste theory down to
one taste with the standard rooting trick
\begin{equation}
  |D|\longrightarrow |D^4|^{1/4}
\end{equation}
is indeed valid.  Once the cutoff is in place, $D$ is a finite matrix
and this a mathematical identity.\footnote{This technically requires
  $|D|$ real and non-negative.  Interesting cases where this is not
  true are not considered here.\cite{Creutz:2013xfa}} Ignoring
doublers, which have been made heavy by the Wilson term, the quark
propagator has a single low-energy pole per species.  On varying the
replica factor $N_t$ away from 4, the state degeneracies evolve as
\begin{equation}
  \begin{matrix}
16&\rightarrow &N_t^2 \longrightarrow_{N_t\rightarrow 1}& 1 \cr
15&\rightarrow& N_t^2-1 \longrightarrow_{N_t\rightarrow 1}& 0 \cr
\end{matrix}
  \end{equation}
and we obtain the proper spectrum.

\section{Staggered fermions}\label{sec-4}

Now turn to the case of staggered fermions.  In this theory, the extra
species are not replicas.  The doublers all appear in chiral pairs.
Whether one roots or not, the propagator always has four light poles.
This means that the spurious 15 multiplets will remain in the
spectrum even after rooting.
Furthermore, the exact chiral symmetry of the staggered approach
requires at least one member of each of these taste-15 multiplets to
become a Goldstone boson in the chiral limit.  Even if we allow for
taste breaking, some remnants of the spurious multiplets must remain.

So if the approach is fundamentally flawed, why do previous
calculations with this method frequently appear to be fairly accurate?
The issues are connected with so called ``disconnected diagrams.''
These are fundamental to the mixing between the strange and the light
quarks inherent in the eta meson.  Most previous staggered
calculations have concentrated on particles dominated by valence
quarks, ones that propagate without such direct mixing.  For these,
the problems are swept into the sea quarks.  Such are known to
contribute of order ten percent relative to results in the
valence\cite{Weingarten:1980hx} or quenched\cite{Hamber:1981zn}
approximation, where the sea is ignored.  Furthermore, the sea quark
contributions will primarily differ because of incorrect
multiplicities in the ``pion cloud.''  In the isospin limit the
staggered theory has 63 degenerate pions.  This is reduced to $63/16$
effective pions after the rooting trick.  This compares with the
physical cloud composed of of 3 pions.  Thus the final error for
valence physics is expected to be reduced to a few percent.  Larger
problems are expected when disconnected diagrams are crucial.  This
should be particularly serious for the physics of the eta and
eta-prime mesons as well as for isospin breaking effects.

\section{Summary}\label{sec-5}

This discussion raises two issues that practitioners of
rooted-staggered fermions should address
\begin{enumerate}
  \item How can the differing taste multiplicities of
    pseudo-scalar-octet members be reconciled?
  \item How can the three unphysical taste-15 multiplets with
    unphysical masses be eliminated?
\end{enumerate}
Without answers to these questions, the approach can at best be
regarded as an uncontrolled approximation to QCD.




\begin{thebibliography}{12}

\bibitem{Bernard:2006qt}
C.~Bernard, M.~Golterman, Y.~Shamir, PoS \textbf{LAT2006}, 205 (2006),
  \texttt{hep-lat/0610003}

\bibitem{Kogut:1974ag}
J.B. Kogut, L.~Susskind, Phys.Rev. \textbf{D11}, 395 (1975)

\bibitem{Susskind:1976jm}
L.~Susskind, Phys.Rev. \textbf{D16}, 3031 (1977)

\bibitem{Sharatchandra:1981si}
H.~Sharatchandra, H.~Thun, P.~Weisz, Nucl.Phys. \textbf{B192}, 205 (1981)

\bibitem{Creutz:2007yg}
M.~Creutz, Phys. Lett. \textbf{B649}, 230 (2007), \texttt{hep-lat/0701018}

\bibitem{Creutz:2007rk}
M.~Creutz, PoS \textbf{LAT2007}, 007 (2007), \texttt{0708.1295}

\bibitem{Wilson:1975id}
K.G. Wilson, Erice Lectures 1975  (1977), new Phenomena In Subnuclear Physics.
  Part A. Proceedings of the First Half of the 1975 International School of
  Subnuclear Physics, Erice, Sicily, July 11 - August 1, 1975, ed. A.~Zichichi,
  Plenum Press, New York, 1977, p.~69, CLNS-321

\bibitem{Lee:1999zxa}
W.J. Lee, S.R. Sharpe, Phys. Rev. \textbf{D60}, 114503 (1999),
  \texttt{hep-lat/9905023}

\bibitem{Aubin:2003mg}
C.~Aubin, C.~Bernard, Phys. Rev. \textbf{D68}, 034014 (2003),
  \texttt{hep-lat/0304014}

\bibitem{Creutz:2013xfa}
M.~Creutz, Annals Phys. \textbf{339}, 560 (2013), \texttt{1306.1245}

\bibitem{Weingarten:1980hx}
D.~Weingarten, D.~Petcher, Phys.Lett. \textbf{B99}, 333 (1981)

\bibitem{Hamber:1981zn}
H.~Hamber, G.~Parisi, Phys. Rev. Lett. \textbf{47}, 1792 (1981), [,619(1981)]

\end{thebibliography}

\end{document}